\documentclass[preprint,12pt]{elsarticle}
\usepackage{amssymb}
\usepackage{amsmath}
\usepackage{graphicx}
\usepackage{booktabs}
\usepackage[unicode]{hyperref}
\usepackage{colortbl}
\usepackage{tabularx}
\usepackage[ruled,vlined]{algorithm2e}
\usepackage{float}
\usepackage{multirow}
\usepackage{array}
\usepackage{makecell}
\usepackage{arydshln}
\usepackage{newtxtext}
\usepackage{newtxmath}
\usepackage{bm}

\journal{Journal of Biomedical Informatics}

\hypersetup{
  pdfauthor={Xinchun Su, Chunxu Luo, Lipeng Ma, Yixuan Li, Weidong Yang},
  pdftitle={MedThink: Enhancing Diagnostic Accuracy in Small Models via Teacher-Guided Reasoning Correction},
  pdfkeywords={Knowledge Distillation, Medicine Reasoning, Large Language Models}
}

\begin{document}

\begin{frontmatter}

%% 标题
\title{MedThink: Enhancing Diagnostic Accuracy in Small Models via Teacher-Guided Reasoning Correction}

\author[Xinchun]{Xinchun Su}
\ead{22300240012@m.fudan.edu.cn}

\author[Chunxu]{Chunxu Luo}
\ead{22110240100@m.fudan.edu.cn}

\author[Lipeng]{Lipeng Ma}
\ead{lpma21@m.fudan.edu.cn}

\author[Yixuan]{Yixuan Li}
\ead{yxli24@m.fudan.edu.cn}

\author[Weidong]{Weidong Yang \texorpdfstring{\corref{cor1}}{}}
\ead{wdyang@fudan.edu.cn}

\address[Xinchun,Chunxu,Lipeng,Yixuan,Weidong]{School of Computer Science, Fudan University, Shanghai, China}

\cortext[cor1]{Corresponding author}

%% 摘要
\begin{abstract}
Accurate clinical diagnosis requires extensive domain knowledge and complex clinical reasoning capabilities. Although large language models (LLMs) hold great potential for clinical reasoning, their high computational and memory requirements limit their deployment in resource-constrained environments. Knowledge distillation (KD) can compress LLM capabilities into smaller models, but traditional KD merely transfers superficial answer patterns and fails to preserve the structured reasoning required for reliable diagnosis.
To address this, we propose a two-stage distillation framework, \textbf{MedThink}, designed to cultivate robust clinical reasoning in small language models (SLMs). In the first stage, a teacher LLM screens data and injects domain-knowledge explanations to fine-tune a student model, establishing a knowledge foundation. In the second stage, the teacher evaluates the student’s errors, generates reasoning chains linking knowledge to correct answers, and refines the student’s diagnostic reasoning through a second round of fine-tuning.
We evaluate MedThink on general medical benchmarks and a gastroenterology dataset comprising 955 question-answer pairs. Experiments demonstrate that MedThink outperforms six distillation strategies in all benchmarks: achieving an improvement of up to 12.7\% over the student baseline in general tasks, and reaching a total top accuracy of 56.4\% in gastroenterology evaluation. This indicates that iterative distillation centered on reasoning can significantly enhance the diagnostic accuracy and generalization capabilities of SLMs whilst maintaining computational efficiency.
Our code and data are publicly available at \url{https://github.com/destinybird/PrecisionBoost}.
\end{abstract}

%% 关键词
\begin{keyword}
Knowledge Distillation \sep Medicine Reasoning \sep Large Language Models
\end{keyword}

\end{frontmatter}

%% 正文
\section{Introduction}
\label{sec:introduction}
Large Language Models, or LLMs, such as GPT-4 and LLaMA perform well on reasoning tasks, including chain-of-thought problem-solving, text generation, and question answering \cite{li2024can, hsieh2023distilling, kojima2022large, wei2022chain}. In medicine, these models can support clinical decision-making, diagnosis, and consultation. However, LLMs require large amounts of GPU memory and have high inference latency, which makes them difficult to deploy in settings with limited resources, such as community clinics, mobile health platforms, and edge devices \cite{tian2024tinyllm, xie2024llms}. This has led to growing interest in Small Language Models, or SLMs, that balance performance with efficiency, making AI-assisted diagnosis more widely accessible.

Knowledge distillation, or KD, offers a way to transfer the capabilities of a large ``teacher'' model to a smaller ``student'' model \cite{li2024can, hsieh2023distilling, ouyang2022training}. Traditional KD methods align model outputs by replicating final answers or distilling chain-of-thought, or CoT, reasoning traces. While these methods can transfer basic knowledge, they are not sufficient for medical tasks that require both domain expertise and sound diagnostic logic \cite{asai2023selfrag, kang2023knowledge}. Accurate clinical diagnosis involves not only recalling symptom-disease associations but also reasoning about disease mechanisms, considering differential diagnoses, and weighing clinical evidence. Standard output-matching distillation does not develop these abilities in student models.

Two key challenges limit existing KD methods in medicine. First, standard distillation lacks \emph{domain-specific knowledge depth}. Student models often fail to correctly interpret medical terminology and clinical context, which reduces diagnostic accuracy \cite{kang2023knowledge}. Second, even when factual knowledge is transferred, student models have difficulty applying it through \emph{coherent reasoning}. They struggle to connect symptom patterns to candidate diagnoses while considering risk factors and disease mechanisms \cite{luo2024step, shen2024self}. These challenges highlight the need for distillation methods that build both domain knowledge and structured reasoning in small models.

To address these challenges, we propose \textbf{MedThink}, a two-stage knowledge distillation framework that enhances diagnostic accuracy in small models through teacher-guided reasoning correction. In the first stage, called \emph{Knowledge Acquisition}, a teacher LLM filters training data by quality and adds domain-specific knowledge explanations to each sample. These enriched samples are used to fine-tune the student model and build a solid medical knowledge base. In the second stage, called \emph{Reasoning Enhancement}, the teacher evaluates the student's errors, identifies where the reasoning went wrong, and generates corrective reasoning chains that connect domain knowledge to the correct answers. The student is then fine-tuned again using these corrective chains to improve its diagnostic reasoning. This two-stage design separates knowledge building from reasoning correction, following the natural process of clinical training, where students first learn the knowledge and then learn how to apply it.

We evaluate MedThink on general medical benchmarks, including CMExam, MedTiku, and ChatMed, and a specialized digestive-domain dataset. MedThink outperforms six alternative distillation strategies across all benchmarks, with up to 12.7\% improvement over the student baseline. In the digestive domain, MedThink achieves the highest overall accuracy of 56.4\%, compared to 53.8\% for the best alternative. These results show that teacher-guided reasoning correction can significantly improve both diagnostic accuracy and generalization in small models, providing a practical approach for deploying AI-assisted diagnostic systems in resource-limited healthcare settings.

The primary contributions of this work are as follows.
\begin{itemize}
\item We propose MedThink, a two-stage knowledge distillation framework that separates knowledge acquisition from reasoning correction, enabling small models to achieve strong performance on medical diagnostic tasks.
\item We introduce teacher-guided mechanisms for each stage, including quality-filtered knowledge augmentation to build domain knowledge and error-driven corrective reasoning to improve diagnostic logic.
\item We conduct experiments on multiple medical datasets, showing that MedThink improves diagnostic accuracy and reasoning quality over existing KD methods while remaining efficient enough for resource-limited deployment.
\end{itemize}

The remainder of this paper is organized as follows. Section~\ref{sec:related_work} reviews related work on LLMs in medicine and knowledge distillation. Section~\ref{sec:method} describes the proposed two-stage framework. Section~\ref{sec:experiments} presents the experimental setup and results. Section~\ref{sec:conclusion} concludes the paper and discusses future directions.

\section{Related Work}
\label{sec:related_work}

\subsection{Large Language Models in Medicine}

Large language models have been widely applied to medical tasks such as clinical decision-making, diagnosis, and medical question answering \cite{chowdhery2022palm, anil2023palm2}. In the general domain, these models benefit from large-scale pretraining and can adapt to new tasks through prompting strategies \cite{tay2023ul2, zhang2023prompting}. In the medical domain, models trained on biomedical text have been applied to named entity recognition, relation extraction, and clinical document classification \cite{pal2023medhalt}. In the Chinese medical domain, several models have been developed for diagnosis and patient consultation. Huatuo, Zhongjing, Bianque, and ChatMed are representative models that fine-tune pretrained foundations on Chinese clinical data \cite{wang2024halt,huatuo2023,yang2024zhongjing,chen2023bianque,zhu2023chatmed}.

To evaluate the performance of these medical models, researchers have introduced specialized benchmarks. MedBench is a large-scale benchmark covering medical report generation and clinical dialogue \cite{cai2023medbench}. CMExam tests whether models can interpret medical descriptions and patient symptoms to produce accurate diagnoses \cite{liu2023cmexam}. We use both benchmarks in our experiments. Although large models perform well on these benchmarks, they require substantial computational resources for training and inference \cite{tian2024tinyllm, xie2024llms}. This limits their practical use in resource-constrained settings such as community clinics and mobile health platforms. As a result, there is growing interest in building smaller models that retain the diagnostic capabilities of larger ones while being efficient enough for real-world deployment.

\subsection{Knowledge Distillation in Medicine}

Knowledge distillation trains a smaller student model to replicate the behavior of a larger teacher model \cite{mcdonald2024trace}. This approach enables the deployment of capable models in settings where hardware or latency constraints prevent the use of full-scale systems.
Recent work has extended knowledge distillation to transfer not only final answers but also reasoning processes. Chain-of-thought distillation transfers step-by-step reasoning traces from teacher models to student models, helping students learn structured problem-solving patterns \cite{hsieh2023distilling}. Self-distillation methods allow models to improve by learning from their own outputs, while multi-stage distillation progressively refines student capabilities through multiple training rounds \cite{xu2024reasoning, tunstall2023zephyr}. Preference-based alignment methods have also been combined with distillation to improve reasoning quality. These approaches have shown promising results on general reasoning benchmarks.

However, existing reasoning distillation methods face important limitations in the medical domain. Most methods transfer reasoning patterns from teacher outputs without checking whether the student has acquired the domain knowledge needed to apply those patterns correctly. When a student model makes diagnostic errors, current methods do not identify whether the error comes from missing medical knowledge or from flawed reasoning logic. Without this distinction, the correction process cannot target the actual source of failure. This gap motivates our proposed MedThink framework, which separates knowledge acquisition from reasoning correction. MedThink uses teacher-guided reasoning correction to first build domain knowledge in the student model, and then specifically address reasoning failures through corrective chains generated by the teacher.

\section{Method}
\label{sec:method}
Accurate clinical diagnosis depends on two abilities. The first is domain knowledge, including disease mechanisms, drug interactions, and symptom-disease associations. The second is structured reasoning, which allows a model to connect clinical evidence to candidate diagnoses and reach a correct conclusion. Traditional knowledge distillation methods transfer final answers or output distributions from a large teacher model to a smaller student model \cite{jiang2023mistral, dubey2024llama}. However, these methods do not teach the student how to reason through a clinical problem step by step. As a result, distilled student models often produce answers that lack medical justification and fail when faced with complex diagnostic scenarios \cite{luo2024step, shen2024self}.

This limitation is particularly serious in the medical domain. Unlike general-purpose tasks, clinical reasoning requires a model to evaluate multiple competing hypotheses, weigh supporting and contradicting evidence, and follow a structured path from symptoms to diagnosis. A student model that memorizes answer patterns without understanding the underlying clinical logic may produce plausible-sounding but incorrect diagnoses. Transferring reasoning chains, performing targeted error correction, and providing explicit knowledge explanations are therefore essential for medical knowledge distillation. These elements go beyond what traditional distillation can offer, because they teach the student not just what the answer is, but why it is correct and how to arrive at it through sound clinical reasoning.

To address this gap, we propose \textbf{MedThink}, a two-stage knowledge distillation framework that enhances diagnostic accuracy in small models through teacher-guided reasoning correction.

\subsection{Overview of the Two-Stage Framework}
MedThink has two stages that work together. The first stage focuses on knowledge distillation. In this stage, the teacher model creates training data that includes medical concepts, disease mechanisms, and clinical terms. The student model learns from this data to build a strong medical knowledge base. The second stage focuses on reasoning enhancement. In this stage, the teacher reviews the student's errors and generates step-by-step reasoning chains that show how to connect medical knowledge to the correct diagnosis. The student then learns from these reasoning chains to improve its diagnostic logic. This two-stage design reflects how medical training works in practice. Medical students first learn foundational knowledge and then develop clinical reasoning skills through guided practice \cite{gao2023scaling, wu2024self}.

\subsection{Knowledge Distillation Stage}
\begin{figure}[H]
    \centering
    \includegraphics[width=0.9\columnwidth, keepaspectratio]{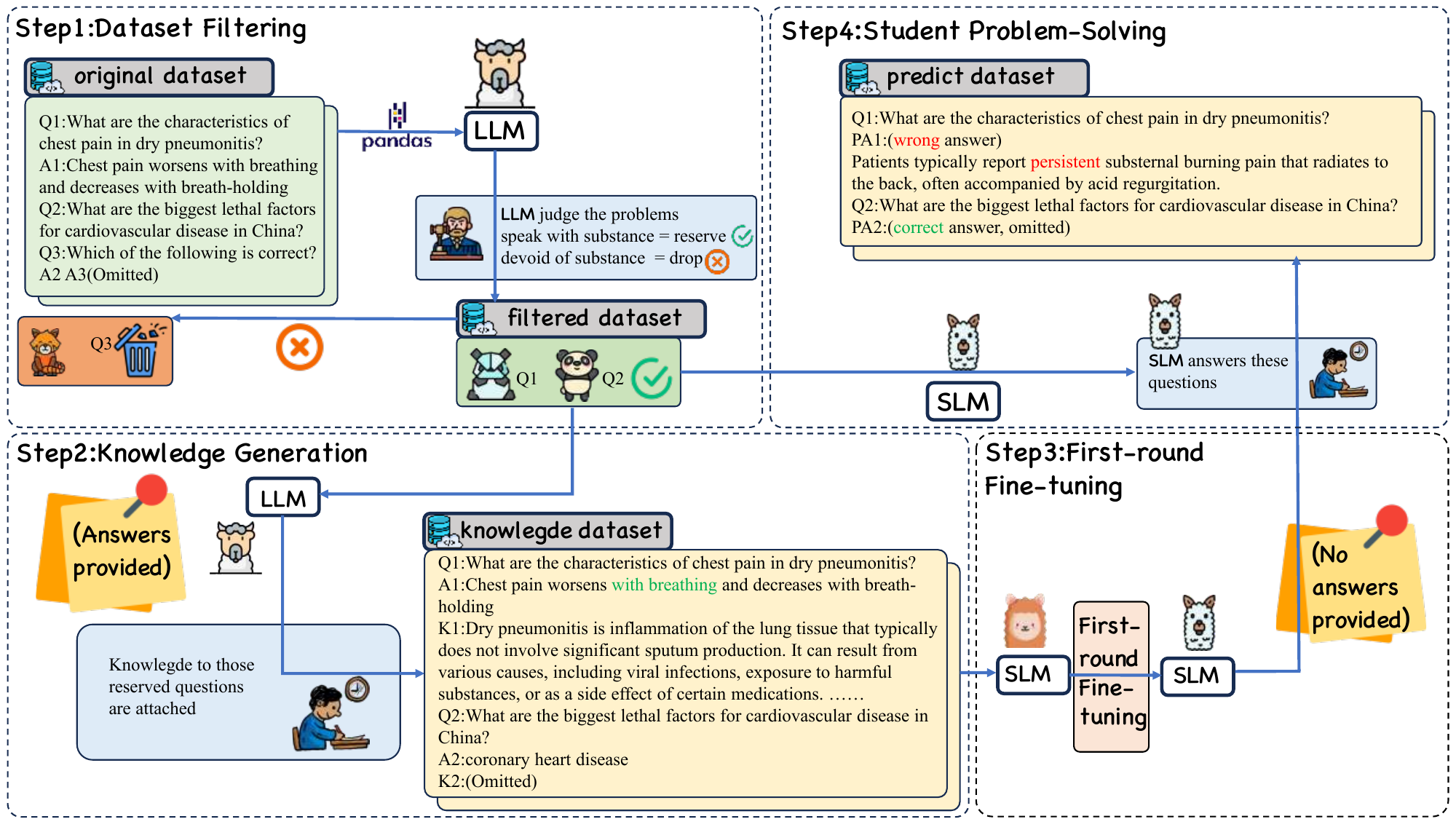}
    \caption{Framework for the Knowledge Distillation Stage.}
    \label{fig:stage1}
\end{figure}

The goal of the first stage is to transfer medical knowledge from the teacher model to the student model. Standard knowledge distillation transfers answers but does not ensure the student understands the medical concepts behind those answers. In MedThink, we use the first stage specifically to build the student's medical knowledge base. This prepares the student for the reasoning training that follows in the second stage. The detailed process is illustrated in Figure~\ref{fig:stage1}.
The process starts with data filtering. The raw dataset \( D \) contains many low-quality samples that could interfere with training. We filter out noisy and ambiguous entries to create a clean dataset \( D_{\text{filtered}} = \{x_i \in D \mid \mathcal{V}(x_i) = 1\} \). Next, the teacher model \( f_{\text{teacher}} \) generates knowledge-augmented triples \( (q_i, a_i, e_i) \) for each question. Here \( q_i \) is the medical question, \( a_i = f_{\text{teacher}}(q_i) \) is the answer, and \( e_i = \Phi(q_i, a_i) \) is an explanation that describes the relevant medical knowledge. These triples form a knowledge base \( K = \{(q_i, a_i, e_i)\} \). The explanations are important because they give the student access to domain knowledge that goes beyond simple question-answer pairs. Unlike traditional distillation that only transfers final outputs, these explicit knowledge explanations teach the student the medical concepts needed for accurate diagnosis. The student model \( f_{\text{student}} \) is then fine-tuned on \( K \) by minimizing the loss \( \mathcal{L}_{\text{round1}} = \frac{1}{|K|} \sum_{i=1}^{|K|} \mathcal{D}(a_i, f_{\text{student}}(q_i; \theta)) \). After this first round of fine-tuning, the student generates predictions \( P = \{(q'_i, \hat{a}_i)\} \) on the training questions. These predictions are used in the second stage to identify where the student still makes errors~\cite{gao2023scaling}.

\subsection{Reasoning Enhancement Stage}
\begin{figure}[H]
    \centering
    \includegraphics[width=\columnwidth, keepaspectratio]{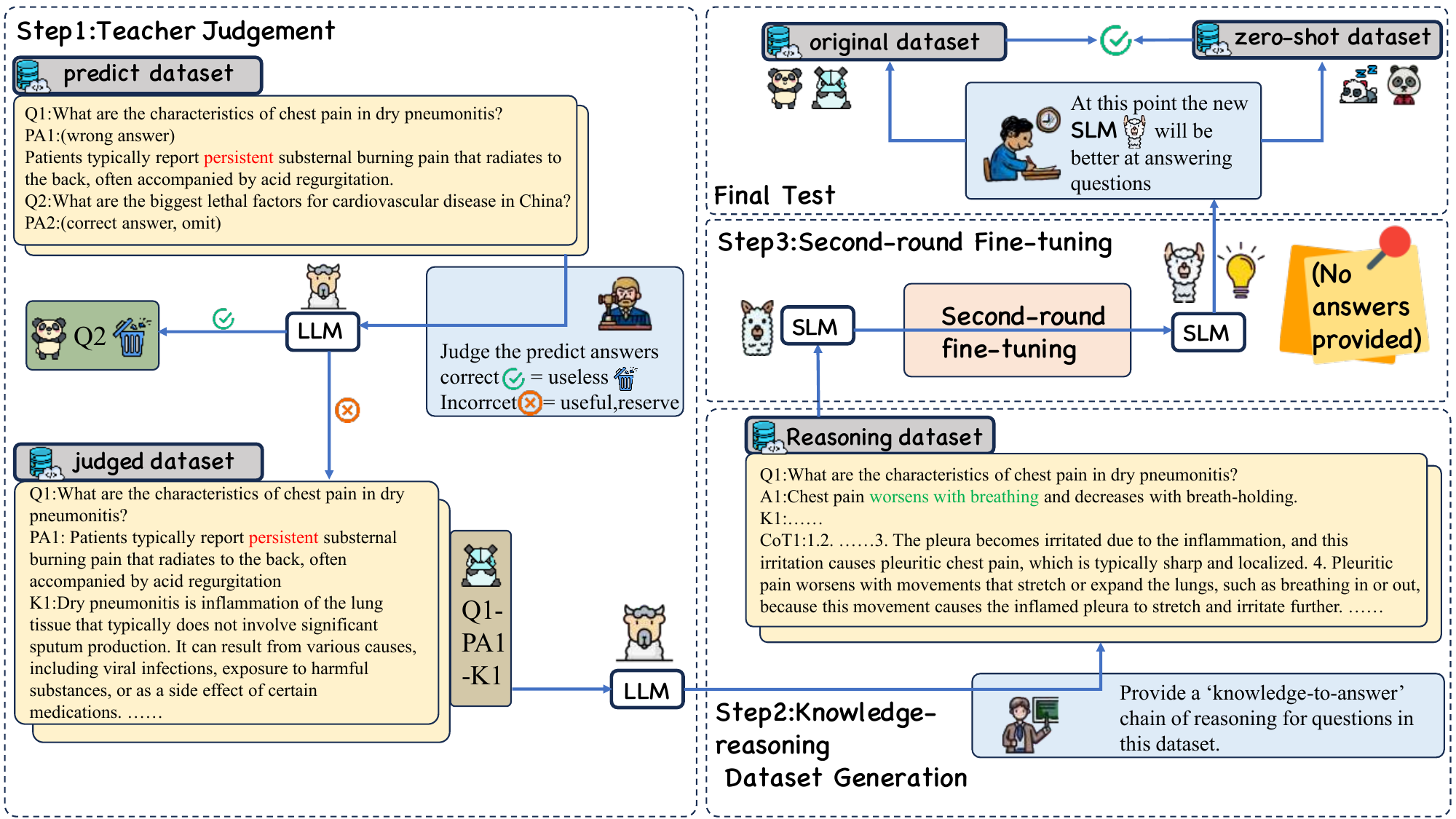}
    \caption{Architecture of the Reasoning Enhancement Stage.}
    \label{fig:stage2}
\end{figure}

After the first stage, the student model has acquired medical knowledge but may still struggle with complex diagnostic reasoning. Small models often fail to connect multiple pieces of evidence in the correct logical order. The second stage of MedThink addresses this problem by teaching the student how to reason through diagnostic problems step by step. The teacher identifies the student's errors and generates corrective reasoning chains that show the correct path from evidence to diagnosis. This targeted error correction is a key advantage over traditional distillation methods, which do not analyze student failures or provide structured guidance for improvement. The process is depicted in Figure~\ref{fig:stage2}.

The teacher model \( f_{\text{teacher}} \) first evaluates the student's predictions \( P \) from the first stage and identifies incorrect outputs \( P_{\text{err}} = \{(q_i, \hat{a}_i) \mid \hat{a}_i \neq f_{\text{teacher}}(q_i)\} \). For each error, the teacher generates a corrective reasoning chain \( c_i \) that explains why the student's answer was wrong and how to reach the correct answer. These form enhanced quadruples \( T = \{(q_i, a_i, e_i, c_i)\} \), where \( a_i = f_{\text{teacher}}(q_i) \) is the correct answer, \( e_i = \Phi(q_i, a_i) \) is the medical knowledge explanation, and \( c_i = \Psi(q_i, a_i, e_i) \) is the corrective reasoning chain. This dataset connects factual knowledge to structured reasoning in an explicit way. Traditional distillation methods do not provide this type of targeted feedback. They transfer the same information to all training samples regardless of whether the student answered correctly or not. In contrast, MedThink focuses the second stage specifically on the questions the student got wrong, making the training more efficient and targeted. The student model is then fine-tuned again on \( T \) by minimizing \( \mathcal{L}_{\text{round2}} = \frac{1}{|T|} \sum_{i=1}^{|T|} \mathcal{D}((a_i, e_i, c_i), f_{\text{student}}(q_i; \theta)) \). Through this second round of training, the student learns to select the best answer \( a_{\text{student}} = \arg\max_{a \in \mathcal{A}} P(a \mid q, e, c; \theta_{\text{student}}) \) by combining medical knowledge with step-by-step reasoning. This error-driven approach is especially valuable in medicine, where understanding why an answer is wrong is just as important as knowing the correct answer~\cite{wu2024self}.

\subsection{Summary of the Workflow}
The following pseudocode summarizes the full workflow of MedThink, from data filtering through both stages of fine-tuning.

\begin{algorithm}[H]
\scriptsize
\caption{Pseudocode of Fine-Tuning Workflow for the Student}
\label{alg:fine_tuning}
\KwIn{Dataset $D$, Teacher Model $f_{\text{teacher}}$, Student Model $f_{\text{student}}$}
\KwOut{Fine-tuned Student Model $f_{\text{student}}$}

\textbf{Step 1. Dataset Filtering} \\
$D_{\text{filtered}} \gets \emptyset$\;
\ForEach{$x \in D$}{
    \If{\text{valid}$(x)$}{Add $x$ to $D_{\text{filtered}}$\;}
}

\textbf{Step 2. Generate Triples} \\
$K \gets \emptyset$\;
\ForEach{$q \in D_{\text{filtered}}$}{
    $a \gets f_{\text{teacher}}(q)$; $e \gets \text{Explain}(q, a)$\;
    Add $(q, a, e)$ to $K$\;
}

\textbf{Step 3. First Fine-Tuning} \\
\ForEach{$(q, a, e) \in K$}{
    Update $f_{\text{student}}$ to minimize $\text{Loss}(a, f_{\text{student}}(q))$\;
}

\textbf{Step 4. Student Prediction} \\
$P \gets \emptyset$\;
\ForEach{$q \in D_{\text{filtered}}$}{
    Add $(q, f_{\text{student}}(q))$ to $P$\;
}

\textbf{Step 5. Teacher Grading} \\
$P_{\text{wrong}} \gets \emptyset$\;
\ForEach{$(q, r) \in P$}{
    \If{$r \neq f_{\text{teacher}}(q)$}{Add $(q, r)$ to $P_{\text{wrong}}$\;}
}

\textbf{Step 6. Generate Quadruples} \\
$T \gets \emptyset$\;
\ForEach{$(q, r) \in P_{\text{wrong}}$}{
    $a \gets f_{\text{teacher}}(q)$; $e \gets \text{Explain}(q, a)$\;
    Add $(q, a, e, \text{ContextExplain}(q, a, e))$ to $T$\;
}

\textbf{Step 7. Second Fine-Tuning} \\
\ForEach{$(q, a, e, c) \in T$}{
    Update $f_{\text{student}}$ to minimize $\text{Loss}(a, f_{\text{student}}(q))$\;
}

\Return{$f_{\text{student}}$}
\end{algorithm}

\section{Experiments}
\label{sec:experiments}
This section presents two experiments to evaluate MedThink. The first experiment tests the general medical domain, where models must generalize across multiple medical subdomains. The second experiment focuses on the specialized digestive domain, where models must handle disease-specific reasoning \cite{ethayarajh2024kto, zhao2024group, chen2024mixed}.

Medical knowledge spans many specialties, each with its own terminology and reasoning patterns. A model trained on general medical data should work well across subdomains but may struggle with topics that need deep expertise. In contrast, a model trained on a specific field like digestive health may perform well in that area but fail to generalize. The general medical experiment evaluates how well MedThink helps models generalize across medical topics. The digestive experiment tests whether MedThink can improve reasoning in a specialized domain.

\subsection{Datasets}
We use several datasets to evaluate MedThink across general medical and specialized digestive health domains.

For the general medical domain, we use three test datasets.
\begin{itemize}
    \item \textbf{CMExam Dataset.} Derived from \url{https://github.com/williamliujl/CMExam/blob/main/data/train.csv}, it contains 68,000 multiple-choice medical questions. After filtration using Kimi’s API, 13,887 high-quality questions were selected. Answer explanations were generated by Qwen2-72B-Instruct.
    \item \textbf{Medtiku Dataset.} Gathered from \url{https://www.medtiku.com/}, it includes 200 question-answer pairs on medical topics. 100 questions were used for training, and 100 for testing.
    \item \textbf{ChatMed Dataset.} Available at \url{https://huggingface.co/datasets/michaelwzhu/ChatMed_Consult_Dataset}, it contains 110,113 doctor-patient consultation pairs generated by GPT-3.5, covering diverse medical inquiries in Chinese.
\end{itemize}

For the specialized digestive domain, we use a dataset with 955 question-answer pairs on digestive diseases.
\begin{itemize}
    \item \textbf{MedQA Dataset.} It is derived from \url{https://github.com/jind11/MedQA}. It is split into 755 training and 200 test entries.
\end{itemize}

These datasets enable broad evaluation across general and specialized medical domains.

\subsection{Baseline}
In our experiments, we use two baseline models. Qwen-7B serves as the baseline for general medical domain tasks, and Qwen2-7.5B-Instruct serves as the baseline for digestive-specific medical tasks. Both models are based on the Transformer architecture and pre-trained on large-scale datasets, primarily Chinese and English texts.

The Qwen-7B model handles a broad range of language tasks, trained on 3TB of diverse data. It uses RoPE for position encoding and Flash Attention for faster processing. For the specialized digestive health domain, we use Qwen2-7.5B-Instruct, fine-tuned with RLHF to improve performance on domain-specific tasks. It uses SwiGLU activation and dynamic NTK perception for extended context length.

These baseline models provide a foundation for evaluating fine-tuning strategies in general and specialized medical domains.

\subsection{Setup}
Specialized medical domains such as digestive health often involve more complex reasoning challenges. These challenges arise from the need for precise domain-specific knowledge and careful diagnostic logic. To evaluate how well MedThink addresses these challenges, we design two experimental scenarios. The general medical domain tests broad applicability across multiple subdomains. The specialized digestive domain tests performance in a focused, complex context.

The setup for both experiments includes preprocessing and knowledge generation steps. In the general medical domain, we first filter the original training dataset using the Kimi API. This filtering step removes about 80\% of the initial data, leaving only high-quality questions for further processing.

After cleaning the dataset, we use the large language model Qwen-72B to generate knowledge for each filtered question. This process produces the knowledge dataset, which consists of question-answer pairs and the knowledge generated by the teacher model. This knowledge dataset serves as the foundation for MedThink, guiding the fine-tuning of the smaller student models in the following steps.

In the specialized digestive domain, the preprocessing and knowledge generation steps are similar, but no filtering is applied. The original dataset, containing 955 question-answer pairs focused on digestive health, is processed directly by the Qwen2-7.5B-Instruct model. This dataset forms the foundation for the training and evaluation of models in the digestive-specific medical domain.

\subsubsection{General Medical Domain}
In this experiment, we apply seven fine-tuning strategies to guide the student model using knowledge distilled from the Qwen-72B model, called Teacher. Each strategy tests a different component of the MedThink framework, from basic knowledge transfer to full reasoning chain integration. The strategies and their corresponding student models are listed below.

\begin{enumerate}
\item \textbf{Strategy 1. OneShot Distill}
The student model is fine-tuned directly using the knowledge from the original dataset. This dataset contains question-answer pairs along with explanations generated by Teacher. The resulting model, OneShot Distill, serves as the baseline. It represents the simplest form of knowledge transfer without reasoning chain refinement or error correction.

\item \textbf{Strategy 2. AnswerFix}
Teacher identifies errors in the output produced by OneShot Distill and corrects them. The student model is then fine-tuned again using only the corrected answers. The resulting model, AnswerFix, tests whether error correction alone can improve reasoning accuracy.

\item \textbf{Strategy 3. CoTCorrection}
Similar to AnswerFix, Teacher identifies and corrects errors in the initial output. However, this strategy also includes explanations of why the incorrect answers were wrong. This additional reasoning helps the student model learn both the correct answer and the logic behind it. The resulting model, CoTCorrection, benefits from both error correction and reasoning chain explanations.

\item \textbf{Strategy 4. DomainMix}
We combine corrected answers and explanations from CoTCorrection with general-purpose data. The DomainMix model tests if broader data improves generalization.

\item \textbf{Strategy 5. IterativeRefine}
An iterative fine-tuning approach refines OneShot Distill with Teacher feedback over two rounds. The IterativeRefine model incorporates multiple feedback cycles.

\item \textbf{Strategy 6. StructureBlend}
The dataset is converted into question-answer-explanation triples, combined with general-purpose data. The StructureBlend model tests structured knowledge integration.

\item \textbf{Strategy 7. MedThink}
This strategy applies the full MedThink framework. Teacher focuses on the questions that OneShot Distill answered incorrectly, generating detailed explanations for each failed case. The student model is then fine-tuned with these targeted corrections and reasoning explanations. The MedThink model combines knowledge explanation, reasoning chain construction, and error correction to refine performance on weak areas \cite{lee2024distilling, wang2023large}.
\end{enumerate}

These seven strategies represent different levels of the MedThink framework, from basic knowledge transfer to full reasoning chain integration with error correction. We compare their performance across test sets for general medical tasks.

\subsubsection{Digestive-Specific Medical Question Answering}
We apply four fine-tuning strategies to the digestive-specific dataset, using knowledge from the Qwen2-7.5B-Instruct model as Teacher. These strategies follow the same MedThink methodology as the general medical domain \cite{liu2024improving, snell2024learning, yuan2024self}.

\begin{enumerate}
\item \textbf{Strategy 1. D-OneShot Distill}
Similar to OneShot Distill, one-round fine-tuning is performed on the training set. The D-OneShot Distill model serves as the baseline.

\item \textbf{Strategy 2. D-AnswerFix}
Similar to AnswerFix, the model is fine-tuned using only question-answer pairs. The D-AnswerFix model tests answer-only fine-tuning.

\item \textbf{Strategy 3. D-StructureBlend}
Similar to StructureBlend, it combines one-round fine-tuning with a second round incorporating correct and incorrect patterns. The D-StructureBlend model enhances generalization.

\item \textbf{Strategy 4. D-MedThink}
This strategy applies the full MedThink framework to the digestive domain. It uses two-round fine-tuning with Teacher's error corrections and reasoning explanations. The D-MedThink model combines targeted error correction with knowledge explanation to improve performance on digestive-specific questions \cite{liu2024improving, snell2024learning, yuan2024self}.
\end{enumerate}

These strategies evaluate how MedThink's iterative feedback and reasoning chain approach performs in the digestive health domain.

\subsection{Evaluation}
The results of both experiments are summarized below. We compare the performance of different fine-tuning strategies and their corresponding models.

For the general medical domain, the performance of different models is presented below.

\begin{figure}[t]
    \centering
    \includegraphics[width=0.9\columnwidth, keepaspectratio]{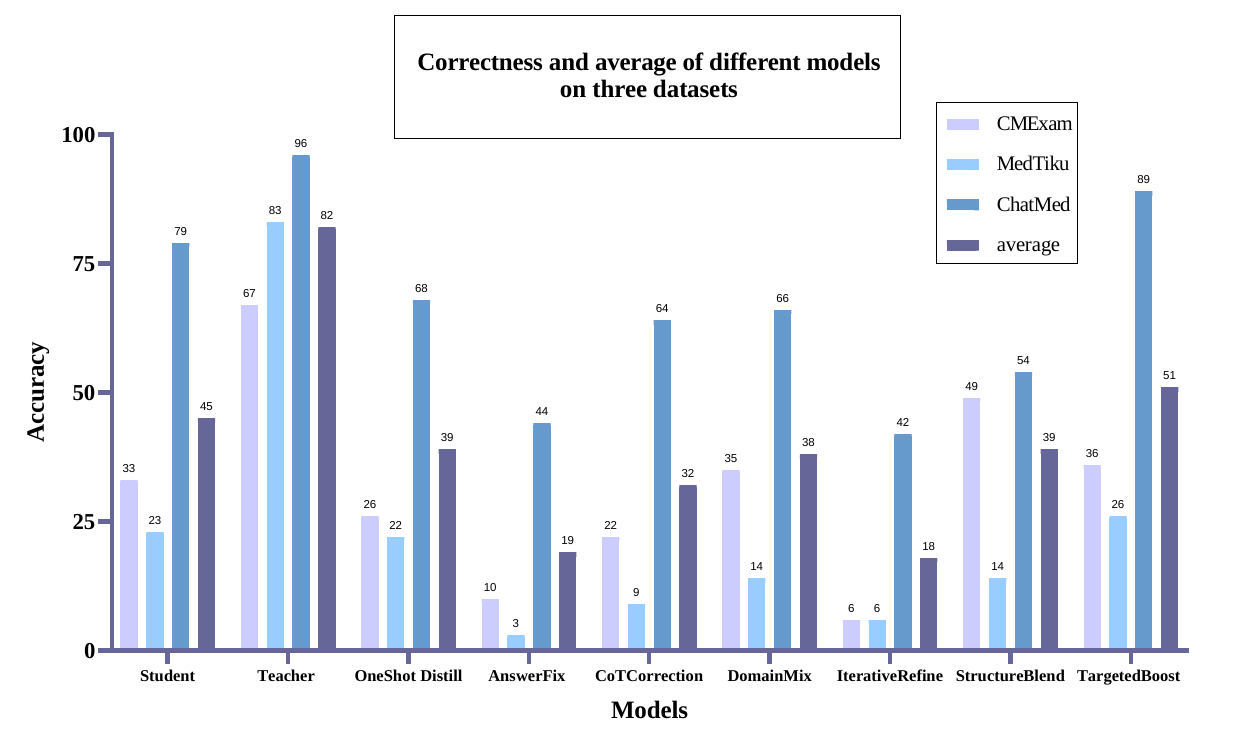}
    \caption{Performance of different models on three test sets, using the first 100 entries, in the general medical domain.}
    \label{fig:results_exp0}
\end{figure}

\begin{table}[H]
\centering
\caption{Model performance in the general medical domain on CMExam, MedTiku, and ChatMed. Scores are accuracy out of 100. MedThink achieves the highest overall scores through two rounds of fine-tuning with knowledge explanation and reasoning chains. Only the first 100 CMExam entries were tested. DomainMix and StructureBlend appear competitive on this subset but remain inferior to MedThink across the full evaluation.}
\label{table:performance_results_exp1}
\begin{tabular}{lcccc}
\toprule
\raisebox{-1.2ex}[0pt][0pt]{ModelName} & \raisebox{-1.2ex}[0pt][0pt]{BaseModel} & \multicolumn{3}{c}{Test Sets} \\
\cmidrule(lr){3-5}
 & & CMExam & MedTiku & ChatMed \\
\midrule
Student & \multirow{2}{*}{None} & 22.4 & 23 & 79 \\
Teacher & & 67 & 83 & 96 \\
\hdashline
OneShot Distill & \multirow{4}{*}{Student} & 26.06 & 22 & 68 \\
AnswerFix & & 10 & 3 & 44 \\
CoTCorrection & & 22 & 9 & 64 \\
DomainMix & & 35 & 14 & 66 \\
IterativeRefine & OneShot Distill & 6 & 6 & 42 \\
StructureBlend & \multirow{2}{*}{Student} & 49 & 14 & 54 \\
\textbf{MedThink} & & \textbf{27.33} & \textbf{26} & \textbf{89} \\
\bottomrule
\end{tabular}
\end{table}

For the specialized digestive domain, the performance of different models is presented below.

\begin{table}[H]
\centering
\caption{Model performance in the specialized digestive domain across training, test, and total sets. Scores show correct answers out of 955 question-answer pairs. D-MedThink achieves the highest total score through two-round fine-tuning with error correction and knowledge explanations.}
\label{table:performance_results_exp2}
\begin{tabular}{lcccc}
\toprule
\raisebox{-1.2ex}[0pt][0pt]{ModelName} & \raisebox{-1.2ex}[0pt][0pt]{BaseModel} & \multicolumn{3}{c}{Test Sets} \\
\cmidrule(lr){3-5}
 & & TRAIN & TEST & total \\
\midrule
DS-Student & NONE & 402 & 112 & 514 \\
\hdashline
D-OneShot Distill & \multirow{4}{*}{DS-Student} & 403 & 118 & 521 \\
D-AnswerFix & & 408 & 107 & 515 \\
D-StructureBlend & & 408 & 122 & 530 \\
\textbf{D-MedThink} & & \textbf{423} & \textbf{116} & \textbf{539} \\
\bottomrule
\end{tabular}
\end{table}

\section{Conclusion}
\label{sec:conclusion}
Our experiments show that answer-level knowledge transfer alone is insufficient for equipping student models with robust clinical diagnostic capabilities. The key lies in guiding student models to construct structured reasoning paths and explainable decision logic. To address this, MedThink adopts a two-stage progressive design. In the first stage, the teacher model filters training data by quality and augments it with domain-specific knowledge explanations, establishing a solid medical knowledge foundation in the student model. In the second stage, the teacher evaluates the student’s prediction errors and generates corrective reasoning chains with causal explanations, which are then used to refine the student’s diagnostic reasoning through a second round of fine-tuning. MedThink achieves excellent accuracy on both general medical benchmarks and a specialized digestive disease task, confirming its adaptability and transferability in clinical scenarios. These findings further indicate that combining knowledge transfer with reasoning enhancement is an effective path toward building high-accuracy and explainable small language models for medicine, offering a practical solution for deploying AI-assisted diagnostic systems in resource-constrained healthcare settings.

\section{Limitations}
\label{sec:limitations}
While our experiments show improved reasoning in smaller models through iterative fine-tuning, our approach has several limitations.
\begin{itemize}
\item \textbf{Explainability.} MedThink's reasoning chains offer some transparency into the model's decisions. However, these explanations may not yet meet the strict interpretability standards required for clinical decision-making.
\item \textbf{Scalability and Generalization.} Our method performs well in controlled settings, but its scalability to larger and noisier datasets remains uncertain. Validating MedThink on multi-department, large-scale clinical data is an important next step.
\item \textbf{Training Data Dependency.} The model's performance, especially in the digestive domain, depends on training data quality. Fine-tuning on limited datasets may not generalize well to unseen data. Data augmentation and cross-domain transfer learning could help address this gap.
\end{itemize}

Addressing these limitations will help make the model more robust and useful in real-world medical settings.

%% 参考文献
\bibliographystyle{elsarticle-num}
\bibliography{references}

\end{document}

\appendix

\section{Appendix}
\label{sec:appendix}

\subsection{Datasets and Model Descriptions}
\label{subsec:datasets_models}
\textbf{Datasets:}  
The datasets used in this study consist of both general-purpose and domain-specific datasets. Their original forms and sources are described below:

\begin{enumerate}
    \item \textbf{General Medical Dataset:}  
This dataset was derived from the \texttt{CMExam} dataset, available at \url{https://github.com/williamliujl/CMExam/blob/main/data/train.csv}. CMExam is a comprehensive repository of over 68,000 multiple-choice questions tailored to test language models for the medical domain. The set includes a question, selected answers, the correct one, and an explanation. Table~\ref{tab:cmexam_statistics} shows the dataset statistics, including sample counts and token counts for questions (\( Q \)), answers (\( A \)), and explanations (\( E \)).

We focused on the initial set of 54,497 training samples, filtering to 13,887 clear and meaningful questions using Kimi's API. Original explanations were replaced with new ones generated by Qwen2-72B-Instruct to avoid confusion.

Example question from CMExam:

ID: 3248  
Question: The trigger of acute exacerbation of heart failure  
Candidate answers:  
A: Infection  
B: Myocarditis  
C: Hypertension  
D: Cardiotoxic Drugs  
E: Myocardial Infarction  
Answer: A  
Explanation: Respiratory tract infection, arrhythmia (e.g., atrial fibrillation), increased blood volume...

\begin{table}[ht]
\centering
\begin{tabular}{lcccc}
\toprule
 & \textbf{Train} & \textbf{Val} & \textbf{Test} & \textbf{Total} \\
\midrule
Questions (Original) & 54,497 & 6,811 & 6,811 & 68,119 \\
Questions (Filtered) & 13,887 & -- & -- & 13,887 \\
Vocabulary Size & 4,545 & 3,620 & 3,599 & 4,629 \\
Max Q Tokens & 676 & 500 & 585 & 676 \\
Max A Tokens & 5 & 5 & 5 & 5 \\
Max E Tokens & 2,999 & 2,678 & 2,680 & 2,999 \\
Avg Q Tokens & 29.78 & 30.07 & 32.63 & 30.83 \\
Avg A Tokens & 1.08 & 1.07 & 1.07 & 1.07 \\
Avg E Tokens & 186.24 & 188.95 & 201.44 & 192.21 \\
Median (Q1, Q3) Q Tokens & 17 (12, 32) & 18 (12, 32) & 18 (12, 37) & 18 (12, 32) \\
Median (Q1, Q3) A Tokens & 1 (1, 1) & 1 (1, 1) & 1 (1, 1) & 1 (1, 1) \\
Median (Q1, Q3) E Tokens & 146 (69, 246) & 143 (65, 247) & 158 (80, 263) & 146 (69, 247) \\
\bottomrule
\end{tabular}
\caption{Dataset statistics for CMExam. Notation: Q=Question, A=Answer, E=Explanation. Filtered data retains only training samples passing Kimi's API quality check.}
\label{tab:cmexam_statistics}
\end{table}

This dataset is described in \cite{liu2023benchmarking}.

Another dataset, \texttt{Q\&A.csv}, was gathered from \url{https://www.medtiku.com/} with 200 samples (100 for training), covering various medical topics. The \texttt{ChatMed.csv} dataset, from \url{https://huggingface.co/datasets/michaelwzhu/ChatMed_Consult_Dataset}, contains 110,113 Chinese doctor-patient query-answer pairs generated by GPT-3.5.

    \item \textbf{Digestive-Specific Dataset:}  
Derived from \texttt{MedQA} at \url{https://github.com/jind11/MedQA}, this dataset includes multiple-choice questions in JSONL format. We used \texttt{train.jsonl} from \texttt{data\_clean/questions/\allowbreak Mainland/4\_options}, filtering 955 questions related to the "Digestive System" (meta\_info: Digestive System). Example:

Question: A 35-year-old female patient presented with fever for over 20 days... Which characteristics of ascites match this patient?  
Options:  
A: SAAG \(> 11\,\mathrm{g/L}\)  
B: White blood cell count does not exceed \(500 \times 10^6/\mathrm{L}\) and lymphocytes predominate  
C: Specific gravity of ascites does not exceed 1.018  
D: Protein content exceeds \(30\,\mathrm{g/L}\)  
E: Mycobacterium tuberculosis culture is mostly positive  
Answer: D  
Meta\_info: Digestive System

\begin{table}[ht]
\centering
\begin{tabular}{lc}
\toprule
\textbf{Meta\_info Category} & \textbf{Number of Questions} \\
\midrule
Past Exam Questions & 6,219 \\
Pediatrics & 3,075 \\
Neurology and Psychiatry & 1,839 \\
Pathology & 1,308 \\
Biochemistry & 1,153 \\
Immunology & 1,099 \\
Musculoskeletal Disorders & 1,095 \\
Microbiology & 1,032 \\
Pharmacology & 1,017 \\
\textbf{Digestive System} & 955 \\
% 其他类别省略
\bottomrule
\textbf{Total} & 24,029 \\
\bottomrule
\end{tabular}
\caption{Breakdown of categories in \texttt{MedQA} (Mainland, 4\_options, train.jsonl). Only the Digestive System category (955 samples) was used.}
\label{tab:medqa_meta_info}
\end{table}

This dataset is detailed in \cite{jin2020disease}.

\end{enumerate}

\textbf{Models:}  
\begin{table}[ht]
\centering
\begin{tabular}{lccc}
\toprule
\textbf{Model Name} & \textbf{Params} & \textbf{Publisher} & \textbf{Release Date} \\
\midrule
Qwen2-72B-Instruct (Teacher) & 72.7B & Alibaba Cloud & June 7, 2024 \\
Qwen-7B (Student) & 7.72B & Alibaba Cloud & August 3, 2023 \\
\bottomrule
\end{tabular}
\caption{Details of the teacher and student models used.}
\label{tab:model_details}
\end{table}

\subsection{Experimental Details and Two-Round Fine-Tuning Process}
\label{subsec:exp_details}
Example question: "Chest pain in oesophageal disease is characterised by?" Answer: "exacerbated by eating".

\begin{enumerate}
    \item \textbf{Stage 1: Knowledge Distillation Stage}  
    The model is trained to acquire foundational knowledge. See Figure~\ref{fig:example1}.

    \begin{figure}[ht]
        \centering
        \includegraphics[width=0.8\columnwidth, keepaspectratio]{example 1.pdf}
        \caption{Stage 1: Knowledge Distillation Process.}
        \label{fig:example1}
    \end{figure}

    \item \textbf{Stage 2: Reasoning Enhancement Stage}  
    The teacher model provides feedback for incorrect predictions, followed by fine-tuning. See Figure~\ref{fig:example2}.

    \begin{figure}[ht]
        \centering
        \includegraphics[width=0.8\columnwidth, keepaspectratio]{example 2.pdf}
        \caption{Stage 2: Reasoning Enhancement Process.}
        \label{fig:example2}
    \end{figure}

    \item \textbf{Final Results and Evaluation}  
    The model’s performance is evaluated. See Figure~\ref{fig:finalresult}.

    \begin{figure}[ht]
        \centering
        \includegraphics[width=0.8\columnwidth, keepaspectratio]{example 3.pdf}
        \caption{Final Results and Model Performance.}
        \label{fig:finalresult}
    \end{figure}
\end{enumerate}

%% 参考文献
\clearpage
\bibliographystyle{elsarticle-num}
\bibliography{references}